\documentclass[runningheads]{llncs}

\usepackage[T1]{fontenc}
\def\doi#1{\href{https://doi.org/\detokenize{#1}}{\url{https://doi.org/\detokenize{#1}}}}
\usepackage{amsfonts}
\usepackage{graphicx}
\usepackage{tabularx}
\usepackage{pifont}
\newcommand{\cmark}{\ding{51}}%
\newcommand{\xmark}{\ding{55}}%
\usepackage{listings}
\usepackage{amsmath}
\usepackage{xcolor}
\usepackage{xspace}
\usepackage[backref=page,colorlinks=true,urlcolor=blue]{hyperref}
\makeatletter
\DeclareRobustCommand\onedot{\futurelet\@let@token\@onedot}
\def\@onedot{\ifx\@let@token.\else.\null\fi\xspace}

\def\ie{\emph{i.e}\onedot}

\makeatother

\newcommand*\samethanks[1][\value{footnote}]{\footnotemark[#1]}

\begin{document}
\title{EchoGNN: Explainable Ejection Fraction Estimation with Graph Neural Networks}
\titlerunning{Explainable Ejection Fraction Estimation with Graph Neural Networks}
%
\author{Masoud Mokhtari\inst{1}\orcidID{0000-0001-9471-5573} \and
Teresa Tsang\inst{2} \and
Purang Abolmaesumi\inst{1}\thanks{Co-Corresponding Authors}\and
Renjie Liao\inst{1}\samethanks}
%
\authorrunning{M. Mokhtari et al.}
%
\institute{Electrical and Computer Engineering, University of British Columbia,
Vancouver, BC, Canada \\
\email{\{masoud, purang, rjliao\}@ece.ubc.ca} \and
Vancouver General Hospital, Vancouver, BC, Canada\\
\email{t.tsang@ubc.ca}}
\maketitle              
\begin{abstract}
Ejection fraction (EF) is a key indicator of cardiac function, allowing identification of patients prone to heart dysfunctions such as heart failure. EF is estimated from cardiac ultrasound videos known as echocardiograms (echo) by manually tracing the left ventricle and estimating its volume on certain frames. These estimations exhibit high inter-observer variability due to the manual process and varying video quality. Such sources of inaccuracy and the need for rapid assessment necessitate reliable and explainable machine learning techniques. In this work, we introduce EchoGNN, a model based on graph neural networks (GNNs) to estimate EF from echo videos. Our model first infers a latent echo-graph from the frames of one or multiple echo cine series. It then estimates weights over nodes and edges of this graph, indicating the importance of individual frames that aid EF estimation. A GNN regressor uses this weighted graph to predict EF. We show, qualitatively and quantitatively, that the learned graph weights provide explainability through identification of critical frames for EF estimation, which can be used to determine when human intervention is required. On EchoNet-Dynamic public EF dataset, EchoGNN achieves EF prediction performance that is on par with state of the art and provides explainability, which is crucial given the high inter-observer variability inherent in this task. Our source code is publicly available at: \url{https://github.com/MasoudMo/echognn}.

\end{abstract}
\section{Introduction}
Ejection fraction (EF) is a ratio indicating the volume of blood pumped by the heart. This measurement is crucial in monitoring cardiovascular health and is a potential indicator of heart failure~\cite{efimportant1,efimportant2}. EF is computed using the stroke volume, which is the blood volume difference in the Left Ventricle (LV) during the End-Systolic (ES) and End-Diastolic (ED) phases of the cardiac cycle denoted by ESV and EDV, respectively \cite{BAMIRA201835}.
These volumes are estimated from ultrasound videos of the heart, \ie echocardiograms (echo), which involves detecting the frames corresponding to ES and ED and tracing the LV region. The manual process of detecting the correct frames and making proper traces is prone to human error. Therefore, the American Society of Echocardiography recommends performing EF estimation for up to 5 cardiac cycles and averaging the results~\cite{LANG20151}. However, this guideline is seldom followed in practice, and a single representative beat is selected for evaluation instead. This results in inter-observer variations from 7.6\% to 13.9\% in the EF ratio~\cite{echonet}. 

Automatic EF estimation techniques aid professionals by adding another layer of verification. Additionally, with the emergence of Point-of-Care Ultrasound (POCUS) imaging devices, which are routinely used by less experienced echo users, automation of clinical measurements such as EF is further needed~\cite{pocus}. However, to be adopted broadly, such automation techniques must be explainable to detect when human intervention is required. Different machine learning (ML) architectures have been proposed to perform automatic EF estimation~\cite{bayesianef,segmentationef,echonet,transformeref,segmentationef2}, most of which lack reliable explainability mechanisms. Some of these models fail to provide the model’s confidence on their predictions~\cite{echonet,transformeref,segmentationef2} or have low accuracy due to unrealistic data augmentation during training and over-reliance on ground truth labels \cite{transformeref}.

In this work, we introduce EchoGNN, a novel deep learning model for explainable EF estimation. Our approach first infers a latent graph between frames of one or multiple echo cine series. It then estimates EF based on this latent graph via Graph Neural Networks (GNNs) \cite{scarselli2008graph}, which are a class of deep learning models that efficiently capture graph data. 
To the best of our knowledge, our work is the first one that investigates GNNs in the context of ultrasound videos and EF estimation.
Moreover, our work brings explainability through latent graph learning, inspiring further work in this domain.
Our contributions are threefold:
\begin{itemize}
    \item[$\bullet$] We introduce EchoGNN, a novel deep learning model for explainable EF estimation through GNN-based latent graph learning.
    \item[$\bullet$] We present a weakly-supervised training pipeline for EF estimation without direct reliance on ground truth ES/ED frame labels.
    \item[$\bullet$] Our model has a much lower number of parameters compared to prior work, significantly reducing computational and memory requirements.
\end{itemize}

\section{Related Work}

Most prior works use Convolutional Neural Networks (CNNs) in their EF estimation pipeline \cite{bayesianef,segmentationef,echonet,transformeref}. Ouyang et al.~\cite{echonet} uses ResNet-based (2+1)D convolutions \cite{resnet} to estimate and average EF for all possible 32-frame clips in an echo, while Kazemi Esfeh et al.~\cite{bayesianef} uses a similar approach under the Bayesian Neural Networks (BNNs) setting. Recent work uses the encoder of ResNetAE \cite{ResNetAE} to reduce data dimensionality before using transformers \cite{posemb} to jointly perform ES/ED frame detection and EF estimation~\cite{transformeref}. While these methods show different levels of accuracy and success in predicting EF, they either lack explainability or significantly rely on accurate clinical labels, which are inherently noisy and subject to significant inter-observer variability. As an example, the transformer-based approach requires ES/ED frame index labels in addition to EF labels in its training pipeline \cite{transformeref}. Lastly, while Kazemi Esfeh et al. and Jafari et al.~\cite{bayesianef,segmentationef} report uncertainty on their predictions, they still lack explainable indicators as to why models fail or succeed for different cases. Our proposed framework based on GNNs aims to alleviate these shortcomings. It provides explainability by only relying on EF labels and not requiring ES/ED frame labels in a supervised manner. Lastly, as an added advantage, the number of parameters for our model is significantly less than prior work, which is highly desirable for deploying such models on mobile clinical devices.

\section{Methodology}
We consider the following supervised problem for EF estimation: assume for each patient $i\in [N]$ in dataset $D$, there is a ground truth EF ratio $y^i \in \mathbb{R}$, and there are K number of echo videos $x^i_{k} \in \mathbb{R}^{T \times W \times H}$, where $k \in [K]$, T is the number of frames, and H and W are the height and width of each frame. The goal of our model is to learn a function $f: \mathbb{R}^{K \times T \times H \times W} \rightarrow \mathbb{R}$ to estimate EF from echo videos. For notational simplicity and since our evaluation dataset only contains one video per patient, we assume that $K=1$. However, it must be noted that our model is flexible in this regard and can handle multiple videos per patient.

\subsection{EchoGNN Architecture}
\begin{figure}
\centering
\includegraphics[width=0.95\textwidth]{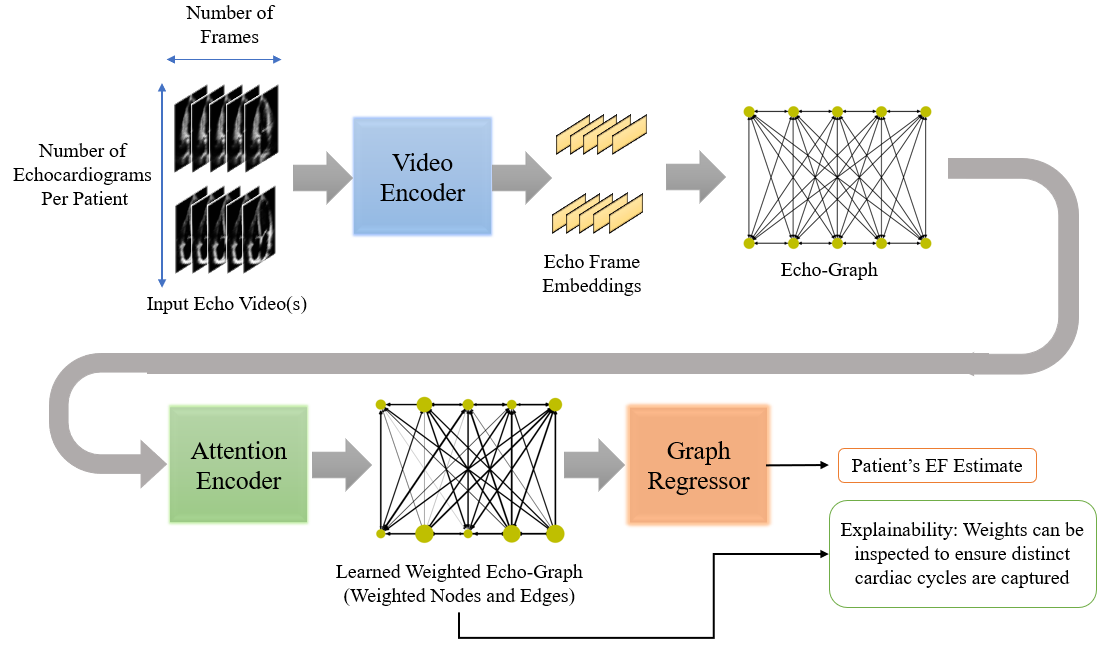}
\caption{EchoGNN has three main components. (1) Video Encoder: encodes video frames into vector embeddings while preserving the temporal dimension; (2) Attention Encoder: infers weights over the nodes (video frames) and edges (relationships among frames) of the echo-graph; (3) Graph Regressor: estimates EF using the inferred weighted graph; this figure shows an example where each patient has an apical two-chamber (AP2) and an apical four-chamber (AP4) echo video.} \label{fig: overall_architecture}
\end{figure}

As shown in Fig.~\ref{fig: overall_architecture}, EchoGNN is composed of three main components: Video Encoder, Attention Encoder, and Graph Regressor. In the following subsections, we discuss the details pertaining to each component.

\subsubsection{Video Encoder}
\label{sec: video_encoder}
The original echo videos are high-dimensional and must be mapped into lower-dimensional embeddings to reduce memory footprint and remove redundant information.

The Video Encoder is used to learn a mapping $f_{\text{ve}}: R^{T \times H \times W} \rightarrow R^{T \times d}$ from input echo videos $x^i \in \mathbb{R}^{T \times W \times H}$ to $d$-dimensional embeddings $h^i_j \in \mathbb{R}^{d}$, where $j \in [T]$ is the frame number. The temporal dimension is preserved because the Attention Encoder requires embeddings for all frames to produce interpretable weights over them. We use a custom network consisting of 3D convolutions and residual connections to use both the spatial and temporal information in the video in generating the embeddings. This network's architecture is provided in the supp. material. Lastly, following~\cite{posemb}, periodic positional encodings are added to the generated frame embeddings to encode the sequential nature of video data.

\subsubsection{Attention Encoder}
\label{sec: attention_encoder}
For each patient, we construct an \textit{echo-graph}, which is a complete graph where each node corresponds to a frame in the echo video, and the edges show the non-Euclidean relationship between these frames. Formally, we denote the echo-graph with $G_{\text{echo}}(V, E)$ where V is the set of nodes corresponding to echo frames such that $|V| = T$, and E is the set of edges between the nodes to show the relationship between video frames such that if $v_1, v_2 \in V$ are connected, then $e_{v_1, v_2} \in E$. We use the frame embeddings from our Video Encoder as node features of $G_{\text{echo}}$. That is, $\{h^i_1, h^i_2, ..., h^i_T\}$ are the set of features for $\{v_1, v_2, ..., v_T\}$. These embeddings can be represented as a matrix $H^i \in \mathbb{R}^{T \times d}$ such that each row is the embedding for a frame in the echo video for patient $i$.

Inspired by~\cite{kipf2018neural}, we propose using GNNs to learn and assign weights to both edges and nodes of the echo-graph. The edge and node weights are learned to encode the importance of each frame (node weights) and the relationships among frames (edge weights) for the final EF estimation.

The Attention Encoder infers weights over edges and nodes of the echo-graph using message passing based GNNs~\cite{messsagepassing}. A single message passing step is enough for each node to capture information from all other nodes due to echo-graph being a complete graph. More specifically, the following operations are used to obtain weights over each edge $e_{v_k,v_s}$:

\begin{align}
    u_{k,s} &= \text{MLP}_1([h^i_k \Vert h^i_s]) \quad & (\text{node} \rightarrow \text{edge}) \\
    v_s & = \text{MLP}_2(\sum\nolimits_{k \neq s} u_{k,s}) \quad & (\text{edge} \rightarrow \text{node}) \\
    z_{k,s} &= \text{MLP}_3([v_k \Vert v_s]) \quad & (\text{node} \rightarrow \text{edge}) \label{eq: edge_weight} \\
    a_{k,s} &= \sigma(z_{k,s}), &
\end{align}
where $\sigma$ is the Sigmoid function, $[. \Vert .]$ is the concatenation operator, and $a_{k,s} \in [0, 1]$ is the inferred weight for the directed edge from $v_k$ to $v_s$. Similarly, weights for each node $w_s \in [0, 1]$ are generated by inserting another $\text{edge} \rightarrow \text{node}$ operation after Eq.~\ref{eq: edge_weight}. All MLPs use two fully connected linear layers with ELU \cite{elu} activation and batch normalization.

\subsubsection{Regressor}
Our Regressor network uses GNN layers with the learned weighted echo-graph to perform EF estimation. Specifically, for each patient, the output of the Attention Encoder can be represented as a weighted adjacency matrix $A \in [0, 1]^{T \times T}$ and a node weight vector $w \in [0, 1]^T$. The Regressor uses $A$ to generate embeddings over frames of the echo video:
\begin{equation}
    H^l = g^l(A, H^{l-1}),\; \; l=1,...,L
\end{equation}
where $H^l \in \mathbb{R}^{T*d_g}$ is the matrix of learned node embeddings at layer l, $H^0$ is the matrix of frame embeddings from the Video Encoder, and $g^l$ is composed of a Graph Convolutional Network (GCN) layer followed by batch normalization and ELU activation~\cite{gcn}. To represent the whole graph with a single vector embedding, the node embeddings are averaged using the frame weights $w$ generated by the Attention Encoder:

\begin{equation}
    h^i_{\text{graph}} = \frac{\sum_{j=1}^T w_j * H^l_{j}}{\sum_{j=1}^T w_j},
\end{equation}
where $H^l_j \in \mathbb{R}^d$ is the $j$th row of $H^l$, and $w_j$ is the $j$th scalar weight in the frame weight vector. $h^i_{\text{graph}}$ is mapped into an EF estimate using an MLP with two fully connected linear layers, ELU activation and batch normalization.

\subsubsection{Learning Algorithm}
The model is differentiable in an end-to-end manner. Therefore, we use gradient descent with Mean-Absolute-Error (MAE) between predicted EF estimates $\tilde{y}^i$ and ground truth EF values $y^i \in Y$ as the optimization objective, which is computed as $L = \frac{1}{N}\sum_{i=1}^N |\tilde{y}^i - y^i|$.

\section{Experiments}
\subsection{Dataset}
We use EchoNet-Dynamic public EF dataset consisting of 10,030 AP4 echo videos obtained between 2016 and 2018 at Stanford University Hospital. Each echo frame has a dimension of $112\times112$, and the dataset provides ESV, EDV, contour tracings of LV, and EF ratios for each patient~\cite{echonet}. We use the provided splits in the dataset from mutually exclusive patients, including 7465 samples for training, 1288 samples for validation, and 1277 samples for testing. The data distribution in the training set is unbalanced with only 12.7\% of samples having EF ratio below 40\%. Clinically, however, such patients are most critical to be detected for timely intervention \cite{efimp,badef}.

\textbf{Frame Sampling}: To stay within reasonable memory requirements, we use a fixed number of frames per echo denoted by $T_{\text{fixed}}$. During training, we uniformly sample an initial frame index $j$ in $[1, T^i_{\text{total}}-T_{\text{fixed}}]$, where $T^i_{\text{total}}$ is the total number of frames in echo video $i$. We then use $T_{\text{fixed}}$ samples starting from $j$. Following \cite{echonet}, we set $T_{\text{fixed}}$ to 64 and use zero padding in the temporal dimension when $T^i_{\text{total}} < T_{\text{fixed}}$. During test time, we extract multiple back to back clips with each clip containing $T_{\text{fixed}}$ frames and the first clip starting from index 0. We use zero padding in the temporal dimension if $T^i_{\text{total}} < T_{\text{fixed}}$ and overlap the last clip with the previous one if the last clip overshoots $T^i_{\text{total}}$. We set $T_{\text{fixed}}$ to 64 and independently estimate EF for each clip and report the average prediction.

\textbf{Data Augmentation}: Occasionally, AP4 echo is zoomed in on the LV region for certain clinical studies \cite{zoom2,zoom1}. To allow learning of this under-represented distribution, we augment our training set by using a fixed cropping window of $90 \times 72$ centered at the top of each frame and interpolating the result to achieve the original $112 \times 112$ dimension, which creates the desired zoom-in effect.

\subsection{Implementation}
\label{sec: imp}
The Video Encoder uses custom convolution blocks with 16, 32, 64, 128, and 256 channels. The Attention Encoder uses a hidden dimension of 128 for MLP layers, and the Regressor uses 3-layer GNN with 128, 64 and 32 hidden dimensions followed by an MLP with a hidden dimension of 16. We use the Adam optimizer \cite{adam} with a learning rate of 1e-4, a batch size of 80, and 2500 training epochs. Our framework is implemented using PyTorch \cite{pytorch} and PyG \cite{pyg}, and the training was performed on two Nvidia Titan V GPUs. 
\textbf{Pretraining:} We use ES/ED index labels in a pretraining step to train the Video Encoder and the Attention Encoder to give higher weights to ES and ED frames.
\textbf{Classification Loss:} We bin the EF values into 4 ranges $[0-30], (30, 40],$ $(40, 55], (55,100]$ and use a cross-entropy loss encouraging the model to learn EF's clinical categories \cite{efimp}.

\subsection{Results and Discussion}
\subsubsection{Explainability}
\label{sec: qual_results}
\begin{figure}
\centering
\includegraphics[width=0.95\textwidth]{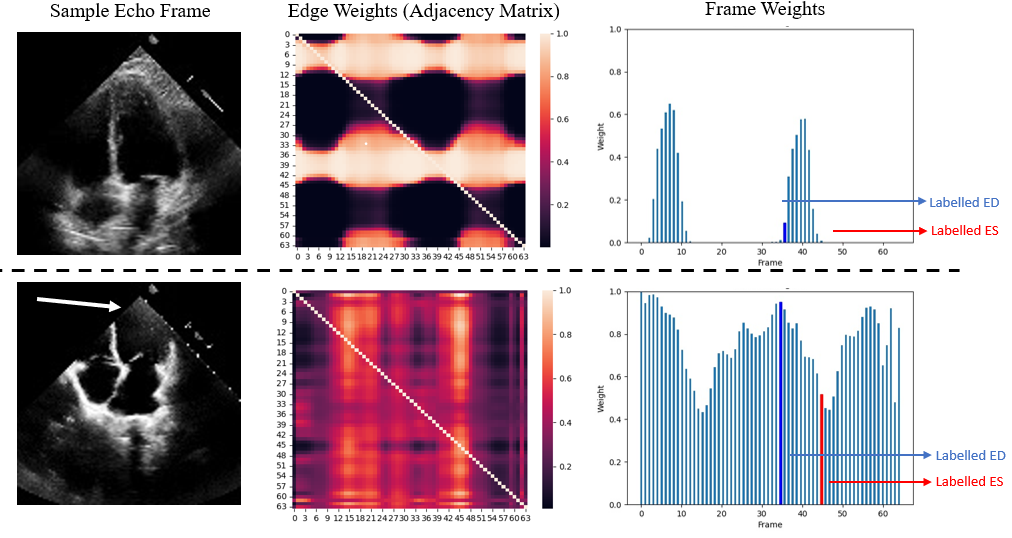}
\caption{(Top) An example where the model has learned the periodic nature of the data, and the learned weights allow identification of ES/ED locations. (Bottom) Another example where the LV region is cropped (as shown by the arrow), and learned weights are distributed more evenly indicating the need for expert intervention.} \label{fig: qual}
\end{figure}

The key advantage of EchoGNN over prior work is the explainability it provides through the learned weights on the echo-graph. As shown in Fig.~\ref{fig: qual}, the learned weights can indicate when human intervention is required. We observe two different scenarios: (1) the model learns the periodic nature of echo videos and assigns larger weights to frames and edges that are in between ES and ED phases before performing EF estimation. This means that the location of ES and ED can be approximated using these weights as illustrated in Fig.~\ref{fig: qual}. (2) The model cannot detect the location of ES and ED frames and distributes weights more evenly. We see that in these cases, we have either an atypical zoomed-in AP4 echo or an echo where the LV is not entirely visible and is cropped. In such cases, an expert can evaluate the video and determine if new videos must be obtained. More explainability examples are provided in the supp. material.

To quantitatively measure the explainability of EchoGNN, for the cases where the model learns the periodic nature of the data (1173 samples out of 1277), we use the average frame distance (aFD) as in \cite{transformeref}, which is computed as $\text{aFD} = \frac{1}{N}\sum_{i=1}^N|j_i-\tilde{j}_i|$ with $j_i$ and $\tilde{j}_i$ being the true and approximated indices, respectively, for sample $i$. As shown in Table \ref{tab1}, our model achieves better ED aFD and comparable ES aFD without using ground-truth ES/ED locations for training, whereas Reynaud et al.~\cite{transformeref} uses such supervision. This shows the explainability power of EchoGNN. aFD computation details are provided in the supp. material.

\subsubsection{EF Estimation}
To evaluate the error in predicted EF values, we use Mean-Absolute-Error (MAE). Additionally, as a measure of the amount of explained variance in the data, we report the model's $\text{R}^2$ score. Moreover, we report the $\text{F}_1$ score for the task of indicating whether EF values are lower than 40\%, which is a strong indicator of heart failure \cite{badef}.

As shown in Table~\ref{tab1}, our model significantly outperforms~\cite{transformeref} without direct supervision of ES and ED frame locations during training. Our model has similar predictive performances as \cite{bayesianef} with a much lower number of parameters and the added benefit of explainability through the learned latent graph structures. EchoNet (AF)~\cite{echonet} requires large amounts of RAM due to sampling all 32-frame clips in a video, making us unable to train and evaluate the model. Because of this we only report results from the paper and cannot produce additional metrics such as $\text{F}_1$ score which is not originally reported, and hence we show this with N/A in Table~\ref{tab1}. This model's weak performance compared to our model shows the sensitivity of EchoNet (AF) to frame locations in a clip. Lastly, our model has a significantly lower number of parameters, making it desirable for deployment on mobile clinical devices. Our model's EF scatter plot and confusion matrix are provided in the supp. material.
\begin{table}
\centering
\caption{Summary of quantitative results. Lower values are better for all metrics besides $\text{R}^2$ and $\text{F}_1$. EchoNet (AF) averages predictions on all possible 32-frame clips in a sampled video. Transformer (R) and (M) are transformer-based models with different sampling techniques. The Bayesian model uses BNNs. 
We mark the models that cannot predict ES/ED locations as "-" in the aFD metric. 
EchoGNN is the only model that provides explainability and ES/ED location estimations without direct supervision.
}\label{tab1}
\begin{tabular}{ l | c>{\centering} c >{\centering}p{0.1\textwidth} |> {\centering}p{0.1\textwidth} >{\centering}p{0.1\textwidth} |>{\centering\arraybackslash}p{0.1\textwidth}}
\hline
Model & $\text{R}^2$ & MAE &  $\text{F}_1$ <40\% & {ES \\aFD} & {ED \\aFD} & \#params  ($\times 10^6)$\tabularnewline
\hline
\hline
EchoNet (AF) \cite{echonet} & 0.4 & 7.35 & N/A & - & - & 31.5 \\
Transformer (R) \cite{transformeref} & 0.48 & 6.76 & 0.70 & \textbf{2.86} & 7.88 & 346.8 \\
Transformer (M) \cite{transformeref} & 0.52 & 5.95 & 0.55 & 3.35 & 7.17 & 346.8 \\
Bayesian \cite{bayesianef} & 0.75 & 4.46 & 0.77 & - & - & 31.5 \\
EchoGNN (ours) & \textbf{0.76} & \textbf{4.45} & \textbf{0.78} & 4.15 & \textbf{3.68} & \textbf{1.7} \\
\hline
\end{tabular}
\end{table}

\subsection{Ablation Study}
\label{sec: ablation}
In Table \ref{tab: tab2}, we see that the classification loss improves model's performance for under-represented samples, while pretraining and data augmentation reduce EF error and increase the model's ability to represent the variance in data.

\begin{table}
\centering
\caption{Ablation study results. Aug., Class., and Pretrain columns indicate if the model uses data augmentation, classification loss and pretraining, respectively. We see that the classification loss improves performance for under-represented groups, while pretraining and data augmentation reduce overall EF error.}\label{tab: tab2}
\begin{tabular}{c >{\centering}p{0.15\textwidth} >{\centering}p{0.15\textwidth} c >{\centering}p{0.15\textwidth} c}
\hline
Aug. & Class. & Pretrain &  $\text{R}^2$ & MAE & $\text{F}_1$ <40\% \\
\hline
\cmark & \xmark & \xmark & 0.75 & 4.48 & 0.76\\
\cmark & \cmark & \xmark & 0.74 & 4.59 & 0.77 \\
\cmark & \xmark & \cmark & 0.75 & 4.47 & 0.73\\
\xmark & \cmark & \cmark & 0.75 & 4.47  & 0.77\\
\cmark & \cmark & \cmark & \textbf{0.76} & \textbf{4.45} & \textbf{0.78}\\
\hline
\end{tabular}
\end{table}

\section{Limitations}
While our model outperforms prior works for EF estimation and also provides explainability, there are certain limitations that can be addressed in future work. Firstly, while the explainability provided over frames and edges of the echo-graph allows identification of cases that need closer inspection, they do not allow finding regions of each frame that the model is uncertain about. We argue that an attention map over the pixels in each frame can further help with explainability. Secondly, creating a complete graph for long videos leads to large memory cost. While this is not an issue for echo, where videos are relatively short, alternative graph construction methods should be considered for longer videos.

\section{Conclusion}
In this work, we introduce a deep learning model that provides the benefit of explainability via GNN-based latent graph learning. While we showcased the success of our framework for EF estimation, we argue that the same pipeline could be used for other datasets and problems, introducing a new paradigm for video processing and prediction tasks from clinical data and beyond.

\subsubsection*{Acknowledgements.} This research was supported in part by the Natural Sciences and Engineering Research Council of Canada (NSERC), the Canadian Institutes of Health Research (CIHR) and computational resources provided by Advanced Research Computing at the University of British Columbia.

\bibliographystyle{splncs04}
\bibliography{ref}

\end{document}


\title{Supplementary Material}
\titlerunning{Explainable Ejection Fraction Estimation with Graph Neural Networks}
%
\author{Masoud Mokhtari et al.}
\authorrunning{M. Mokhtari et al.}
%
\institute{Electrical and Computer Engineering, University of British Columbia,
Vancouver, BC, Canada}
\maketitle              


\begin{figure}
\centering
\includegraphics[width=0.95\textwidth]{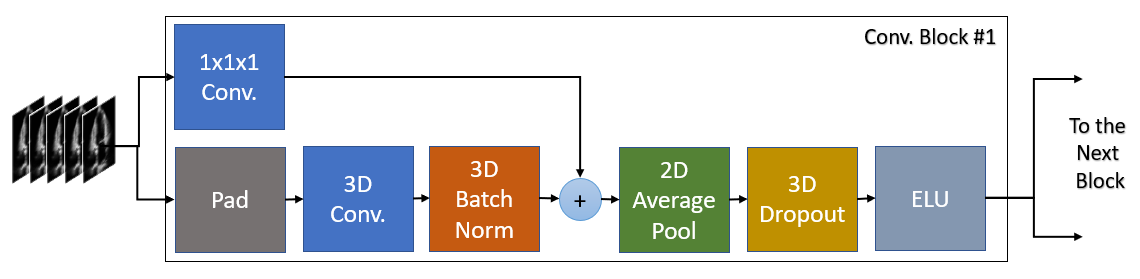}
\caption{Video Encoder network architecture. We use modular blocks containing 3D convolutions with residual connections to generate low-dimensional frame embeddings.} \label{fig: video_encoder}
\end{figure}

\begin{figure}
\centering
\includegraphics[width=1.0\textwidth]{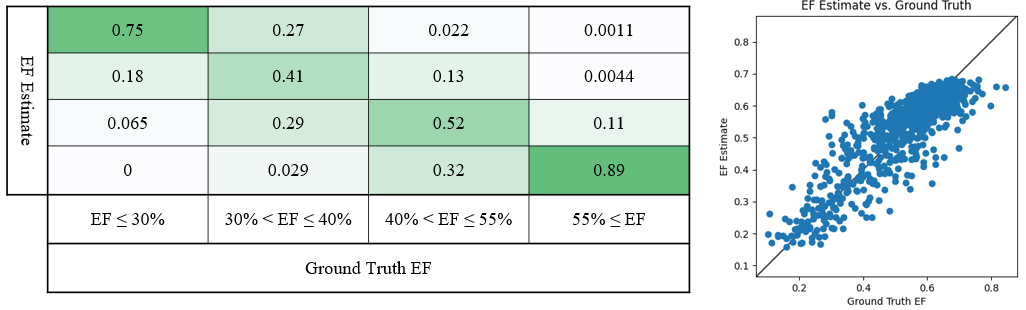}
\caption{(left) The confusion matrix for our best-performing model. The chosen EF categories indicate different levels of heart failure risk with patients having EF below 40\% needing medical monitoring. (right) the scatter plot showing how close our model's EF estimates are to the ground truth. We see that the model struggles with EF values between 30\% and 40\%, and we argue that this is due to the high inter-observer varaibility in the ground truth labels, which is more prominent for samples that lie in pathological boundaries.} \label{fig: scatter}
\end{figure}


\begin{figure}
\centering
\includegraphics[width=1.0\textwidth]{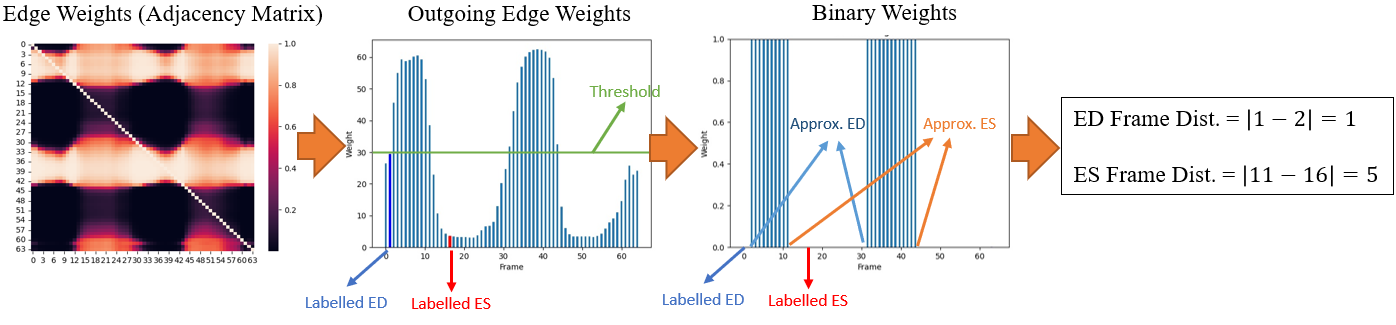}
\caption{ED/ES frame approximation from learned echo-graph weights: we first use a threshold to change the the sum of outgoing edge weights into binary format (alternatively, frame weights can be used). Please note that this threshold is selected based on aFD performance on the validation set. The consecutive 1-valued weights form a block together. The left-most and the right-most frame in each block is the approximated ED and ES locations, respectively. We reject samples where the size of the block is equal to 55, meaning that the model has not learned the periodic nature of data. Rejecting these samples, we achieve an average frame distance of 4.15 for ES and 3.68 for ED.} \label{fig: framedist}
\end{figure}



\begin{figure}
\centering
\includegraphics[width=1.0\textwidth]{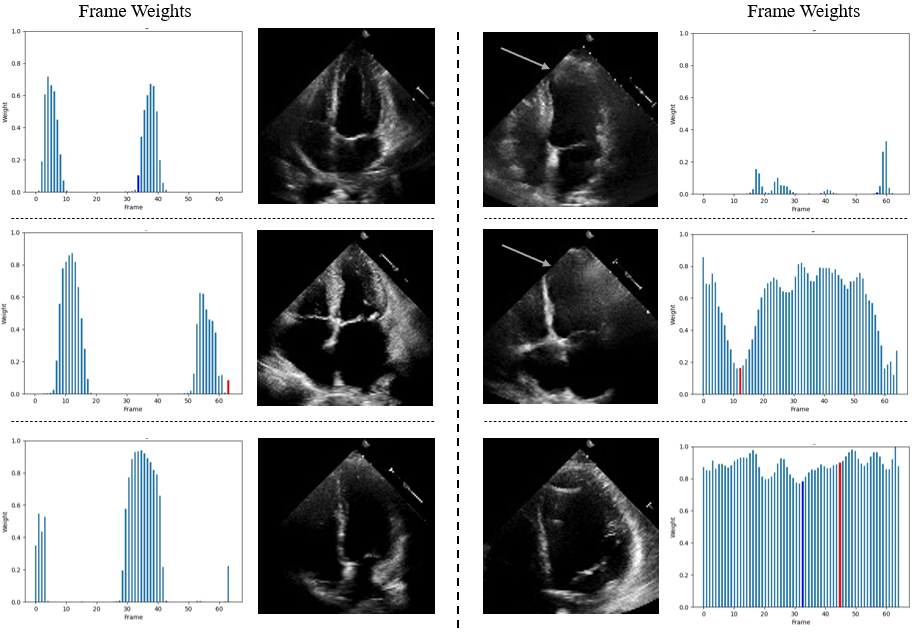}
\caption{Examples of model's explainability capability. (left) We can see examples where the learned frame weights allow clear identification of ES/ED locations. (right) We see examples where we have atypical zoomed-in AP4 echo or echo where the LV is not entirely visible and is cropped, and therefore, the model distributes frame weights more evenly, not clearly indicating the position of ED and ES.} \label{fig: qual}
\end{figure}
